\documentclass{moriond}

\def\be{\begin{equation}}
\def\ee{\end{equation}}
\def\bea{\begin{eqnarray}}
\def\eea{\end{eqnarray}}

\newcommand{\Photo}{}

\usepackage[capitalise]{cleveref}

\begin{document}
\vspace*{4cm}
\title{Interactions in the dark sector: intrinsic entropy couplings}

\author{Elsa M. Teixeira}

\address{Laboratoire Univers \& Particules de Montpellier,\\
CNRS \& Université de Montpellier (UMR-5299), 34095 Montpellier, France}

\maketitle\abstracts{We present a new class of interacting dark sector models in which the intrinsic entropy of a dark matter fluid couples to a scalar field describing dark energy. These interactions are constructed within a covariant Lagrangian framework, including algebraic and derivative entropy couplings, effectively leading to pure momentum exchange in the dark sector. A key feature is that the background cosmology remains unchanged and therefore indistinguishable from $\Lambda$CDM or uncoupled quintessence. However, at the level of the linear perturbations, the dark matter Euler equation exhibits scale-dependent contributions, while the continuity equation is unmodified. We show that these classes of models are compatible with current CMB constraints and can potentially produce observable signatures in large-scale structure.}

\section{Introduction}

The standard $\Lambda$CDM model successfully describes a wide range of cosmological observations, yet the physical nature of dark matter (DM) and dark energy (DE), and their interactions, remain unknown. Moreover, persistent tensions in independent measurements of the Hubble parameter ($H_0$) \cite{Riess:2021jrx,H0DN:2025lyy} and the matter clustering amplitude ($\sigma_8$) \cite{Wright:2025xka,DES:2025tna} motivate the search for resolutions beyond the standard model \cite{CosmoVerseNetwork:2025alb}.
Interactions in the dark sector are expected unless prevented by additional symmetries and provide a natural framework for exploring new physics. However, many existing interaction models either rely on phenomenological constructions lacking a clear theoretical origin or are heavily constrained by modifications to the background expansion history. Recently, models featuring pure momentum exchange have attracted interest within the community as they can affect structure formation while preserving the background evolution. Examples include the Type-3 models of Ref. \cite{Pourtsidou:2013nha}, velocity-entrainment models \cite{BeltranJimenez:2021wbq}, and other phenomenological realisations.

In this work, we introduce a new class of interacting dark sector models in which the coupling to a scalar field dark energy $\phi$ is mediated by the \textit{intrinsic entropy} of the dark matter fluid. These interactions provide a theoretically consistent realisation of pure momentum exchange arising from entropy gradients rather than the previously studied velocity couplings. This theory is constructed within the Lagrangian framework of perfect cosmological fluids developed by Schutz and Sorkin \cite{Schutz:1977df} and Brown \cite{Brown:1992kc} and commonly applied to non-minimally coupled models \cite{Pourtsidou:2013nha,Boehmer:2015sha}, ensuring a consistent variational formulation of the perturbations.
Focusing on an entropic-CDM fluid, which is barotropic and adiabatic with vanishing pressure and sound speed, 
we show that while the background entropy is constant, spatial entropy perturbations represent genuine new primordial degrees of freedom propagated through the initial conditions of the dark sector, affecting the subsequent cosmological evolution.
In particular, the background cosmology remains unchanged with the introduction of entropy couplings, while characteristic scale-dependent signatures arise at the level of perturbations.

\section{Entropy-coupled dark sector and background evolution}

We describe dark matter as a relativistic perfect fluid with intrinsic entropy per particle $s$, using the covariant Lagrangian formulation of Brown. This framework provides a consistent variational description of relativistic perfect fluids with thermodynamic degrees of freedom \cite{Schutz:1977df,Brown:1992kc}, based on the following action:
\begin{equation} 
    S_{\textrm{Brown}} = \int \Big[-\sqrt{-g} \rho(n,s) + \mathcal{N}^\mu \left(\varphi_{,\mu} + s \theta_{,\mu} + \beta_{A}\alpha^{A}_{, \mu} \right) \Big]  \textrm{d}^4x \, . \label{S_Brown}
\end{equation}
Here, $n$ and $\rho$ are the fluid’s particle number density and energy density, $(\varphi,\theta,\beta_A)$ are Lagrange multipliers and $\alpha^A$ label the fluid flow. We define $\mathcal{N}^{\mu} = \sqrt{-g} n u^{\mu}$ as the densitised particle number flux, written in terms of the fluid four-velocity $u^{\mu}$, which satisfies $u^{\mu}u_{\mu}=-1$. 
We include entropy interactions by adding terms of the form
\begin{equation} 
\mathcal{L}_{\rm int} = f(s,\phi,\mathcal{S})\, ,
\end{equation}
where $\mathcal{S}=\nabla_\mu s \nabla^\mu \phi$. This formulation allows both algebraic couplings $g(s,\phi)$ and derivative couplings $h(\nabla_\mu s \nabla^\mu \phi)$ \cite{Boehmer:2015kta,Boehmer:2015sha}.
Entropy-related effects have been neglected in the literature because the background entropy is conserved along the fluid flow, $u^\mu \nabla_\mu s = 0$,
implying that $s$ must be constant at the background level and decouples from the system. 
The interaction reintroduces this degree of freedom into the dynamics,
in a way that does not alter the background expansion history, up to a redefinition of the scalar field potential. Therefore, the background evolution remains indistinguishable from $\Lambda$CDM or uncoupled quintessence. However, as we discuss in the next section, entropy perturbations can still play a significant role in the growth of structures.

\section{Cosmological perturbations}

The entropy perturbations $\delta s$ are non-vanishing and represent additional primordial degrees of freedom. 
A key result of this framework is that the interaction corresponds to a \emph{pure momentum exchange} within the dark sector. This can be understood from the structure of the energy-momentum conservation equations. The projection along the fluid four-velocity vanishes, implying that there is no energy transfer between dark matter and dark energy. As a consequence, the continuity equation remains unchanged at all orders.

In contrast, the spatial projection of the conservation equations leads to a modification of the Euler equation. At linear order and in the Newtonian gauge, this takes the form 
\begin{equation} 
\theta'_c + \mathcal{H}\theta_c = k^2 \Phi - k^2\,Q_s(\tau)\,\delta s (k),
\label{thetac}
\end{equation}
where $\Phi(\tau,\vec{x})$ and $\Psi(\tau, \vec{x})$ are the Newtonian gauge scalar modes, $\theta_c$ is the dark matter velocity divergence, $Q_s(\tau)$ encodes the dependence on the entropy coupling and on the background scalar field evolution
and the scale dependence is determined by the spectrum of $\delta s(k)$.

Entropy perturbations are non-dynamical and act as spatially varying sources, $\delta s=\delta s(\vec{x})$, so that their impact arises entirely through gradients, making the modification to the Euler equation intrinsically scale-dependent.
This resembles Type-3 momentum-exchange models \cite{Pourtsidou:2013nha}, but here the interaction originates from intrinsic entropy degrees of freedom rather than velocity couplings, which can be interpreted as generating effective imperfect fluid properties in the dark sector. In particular, derivative entropy couplings induce heat-flux terms and higher-order corrections to the perfect fluid energy-momentum tensor, which play a role in shaping the perturbation dynamics.

\section{Impact on observables}

In this work we consider the following parameterisation:
\begin{equation} 
\delta s(k) = A_e \left(k/k_p\right)^{n} \exp\left[-\left(k/k_c\right)^{p_c}\right],
\label{eq:deltas_def}
\end{equation}
where $A_e$ is the amplitude, $n$ is the the spectral index, $k_p$ is a pivot scale, $k_c$ defines a cut-off scale that regulates small-scale behaviour, and $p_c$ determines its sharpness.
For simplicity, we focus on the scale-invariant case ($n=0$) and the derivative model $h(\mathcal{S})=h_0 \mathcal{S}$, 
which captures the general phenomenology \cite{Jensko:2026taf}.

In the quasi-static regime, $k \gg aH$, the coupling reduces to an Euler source term $Q_s(\tau)\,\delta s(k)$ together with an effective density contribution to the Poisson equation. In this limit, the effects are negligible at early times and become relevant only at late times, where the Euler term dominates and acts as a fifth-force-like contribution to dark matter dynamics. On large scales ($k \ll aH$), the gradient terms are suppressed, although residual effects can still source the metric potentials through the Einstein equations and contribute to the late-time integrated Sachs-Wolfe effect \cite{Jensko:2026taf}.

\begin{figure}[h]
      \includegraphics[width=0.48\linewidth]{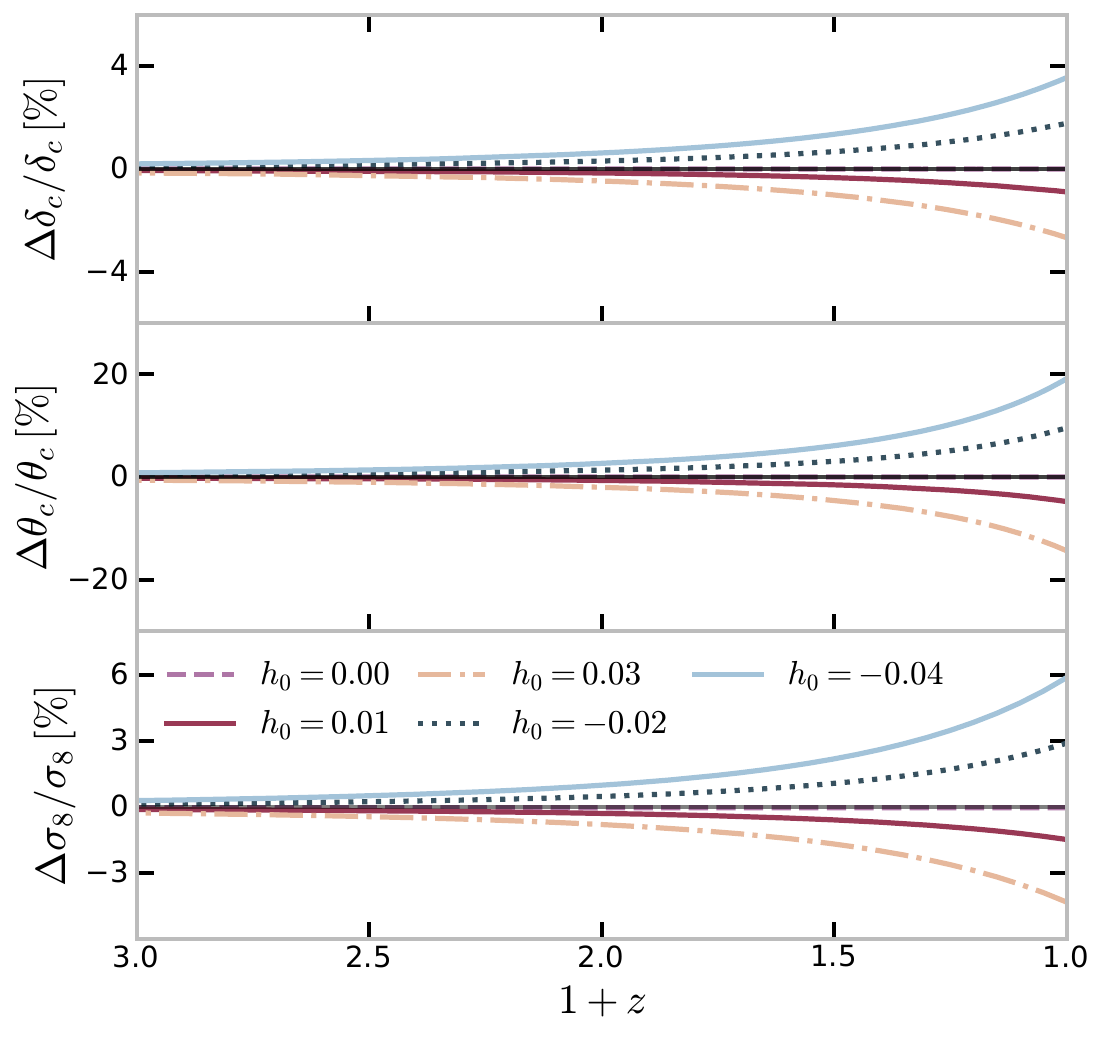}
      \hfill
     \includegraphics[width=0.48\linewidth]{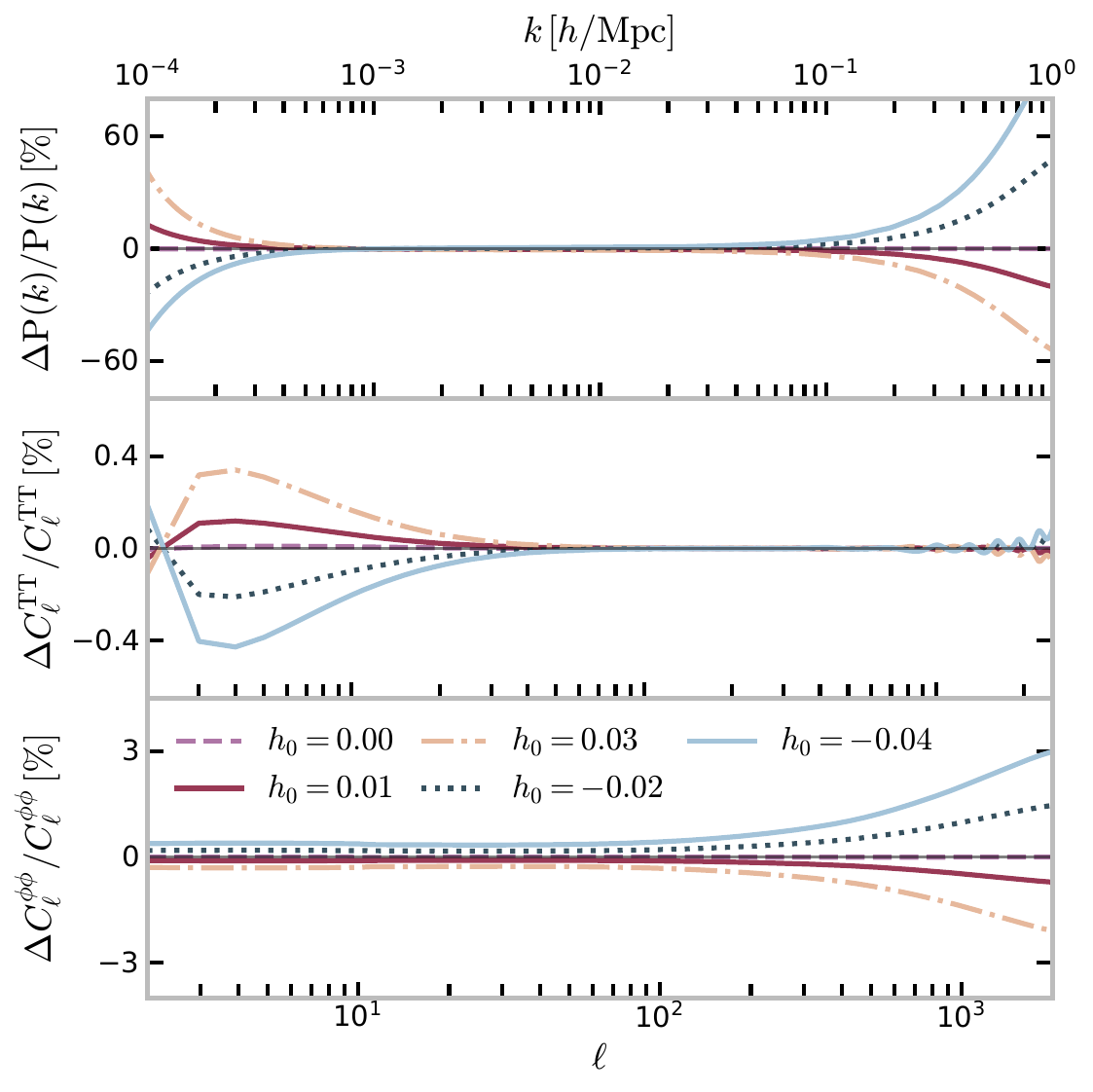}
     \caption{\textit{Left:} Relative deviations of the entropic-CDM perturbations and clustering amplitude with respect to $\Lambda$CDM and as a function of redshift $z$ for $k_8=0.125\,h\,{\rm Mpc}^{-1}$. The top and middle panels show the fractional changes in the dark matter density contrast $\delta_c$ and velocity divergence $\theta_c$, while the bottom panels show the relative change in $\sigma_8$. \textit{Right:} Relative deviations in the matter power spectrum $P(k)$ (top panels), the CMB spectrum of temperature anisotropies $C_\ell^{TT}$ (middle panels), and the CMB lensing potential spectrum $C_\ell^{\phi\phi}$ (bottom panels), with respect to $\Lambda$CDM. In both panels the entropy perturbation $\delta s$ is parameterised as in Eq.~\ref{eq:deltas_def} with $\mathcal{A}_{\rm e}=1$, $n=0$, $k_p=0.05\,\mathrm{Mpc}^{-1}$, $k_c=1\,\mathrm{Mpc}^{-1}$, and $p_c=2$, and we show the derivative coupling case $h(\mathcal{S})=h_0 \mathcal{S}$.}
  \label{fig:delta_theta_DM}
\end{figure}

The evolution of dark matter perturbations and $\sigma_8$ is shown in the left panel of Fig.~\ref{fig:delta_theta_DM}. As expected, the entropy coupling is negligible at early times but becomes significant at late times, where the modification to the velocity divergence leads to an enhancement or suppression of structure growth depending on the sign of the coupling parameter $h_0$. 
The right panel shows the corresponding observable signatures. 
The matter power spectrum exhibits scale-dependent deviations that grow towards small scales. The absence of modifications to the background expansion implies that the CMB temperature anisotropies are only weakly affected, with sub-percent changes at low multipoles consistent with a late-time ISW contribution. These signatures are amplified in the CMB lensing potential, reflecting the modifications to the metric potentials and the cumulative effect of modified structure growth along the line of sight.

\section{Conclusions}

In this work we have presented a new class of interacting dark sector models in which dark matter couples to a scalar field dark energy through its intrinsic entropy. Within a covariant Lagrangian framework, these interactions provide a theoretically consistent realisation of pure momentum exchange, distinct from previously studied velocity-mediated couplings. A key feature of this construction is that the background expansion history is unchanged, while entropy perturbations introduce non-trivial modifications at the level of cosmological perturbations.

We have shown that entropy perturbations act as non-dynamical, spatially varying sources that modify the Euler equation through a scale-dependent term, leading to characteristic imprints in the growth of structure. In particular, the coupling induces a late-time, scale-dependent enhancement or suppression of clustering, while remaining compatible with current CMB constraints.
This makes entropy-coupled models a compelling and testable extension of $\Lambda$CDM, as they evade stringent background constraints while producing distinctive, scale-dependent signatures in structure growth that can be probed with current and future observations.
These results highlight entropy-mediated interactions as a well-motivated and testable extension of $\Lambda$CDM.

\section*{Acknowledgments}

I thank the organisers of the 2026 Cosmology session of the 60th Rencontres de Moriond and the participants for many interesting discussions.
EMT is supported by funding from the European Research Council (ERC) under the European Union's HORIZON-ERC-2022 (grant agreement no. 101076865). 
The results presented in this contribution are based on work carried out in collaboration with Erik Jensko and Vivian Poulin, who also provided helpful comments on this manuscript.

\section*{References}
\bibliography{moriond}

\end{document}